\documentclass[10pt]{article}
\usepackage[letterpaper]{geometry}
\usepackage{hicss}
\usepackage{times}
\usepackage[none]{hyphenat}
\usepackage{url}
\usepackage{latexsym}
\usepackage[frozencache,cachedir=.]{minted}
\usepackage{indentfirst}
\usepackage{graphicx}
\graphicspath{{images/}}
\usepackage[
    style=apa,
  ]{biblatex}
\usepackage[T1]{fontenc}

\usepackage{tabularx}
\usepackage{float}

\title{Simulating Strategic Reasoning: Comparing the Ability of Single LLMs and Multi-Agent Systems to Replicate Human Behavior}

\author{Karthik Sreedhar \\
  Columbia University \\
  {\underline{ks4190@columbia.edu}} \\ \\
   \\
   \\
 {\underline{} } \\ \And
  Lydia Chilton \\
  Columbia University \\
 {\underline{chilton@cs.columbia.edu} } }

\date{}

\begin{document}
\maketitle
\begin{abstract}
When creating policies, plans, or designs for people, it is challenging for designers to foresee all of the ways in which people may reason and behave. Recently, Large Language Models (LLMs) have been shown to be able to simulate human reasoning. We extend this work by measuring LLMs' ability to simulate strategic reasoning in the ultimatum game -- a classic economics bargaining experiment. Experimental evidence shows human strategic reasoning is complex -- people will often choose to ``punish" other players to enforce social norms even at personal expense. We test if LLMs can replicate this behavior in simulation, comparing two structures: single LLMs and multi-agent systems. We compare their abilities to (1) simulate human-like reasoning in the ultimatum game, (2) simulate two player personalities, greedy and fair, and (3) create robust strategies that are logically complete and consistent with personality. Our evaluation shows that multi-agent systems are more accurate than single LLMs (88\% vs. 50\%) in simulating human reasoning and actions for personality pairs. Thus, there is potential to use LLMs to simulate human strategic reasoning to help decision and policy-makers perform preliminary explorations of how people behave in systems.
\end{abstract}

\subsubsection*{Keywords:}

strategic reasoning, large language models, multi-agent systems, social simulation

\section{Introduction}

Simulations help us design the world. 
When creating plans and policies, such as worker paths on factory shop floors \parencite{herr2019bluecollar}, introducing new technologies into a company's workflow \parencite{kasik2019toward}, or displaying several routes on the same map \parencite{zhao2020route},  
simulations help decision-makers think through possible actions and outcomes in complex systems.
Although physical simulation has become highly sophisticated in recent years ~\parencite{ThreeDWorld}, simulating human behavior remains notoriously difficult. When economists model human behavior, they assume that people are rational actors, but psychology has discovered many important cases in which human reasoning is not driven by profit maximization ~\parencite{kahneman_thinking_2012, ariely_2008}. Moreover, people do not reason uniformly -- their personalities~\parencite{mccrae2008five}, experiences~\parencite{KIDD2013109} and circumstances~\parencite{mullainathan2013scarcity} affect their decision making. Strategic reasoning adds another complexity -- in some scenarios, people base their actions off of the demeanor and actions of others (as chess players do). This makes it mentally demanding for decision-makers to foresee all of the possibilities of how people would act in response to new policies.

Recently, LLMs have been shown to be able to simulate human reasoning based on personality types. This includes modeling the opinions of supreme court justices in past rulings ~\parencite{hamilton2023blind}, simulating a fictional town’s ability to plan and attend events like a party ~\parencite{park2023generative}, and simulating human behavior in classic economic and psychology experiments ~\parencite{aher2023using}. We extend this prior work by investigating whether LLMs can simulate human strategic reasoning by comparing LLM simulation results to experimental human baselines.

The ultimatum game is a classic economics experiment used to study human strategic reasoning with social context. 
It captures human social behavior (often deemed irrational, such as the desire to “punish” unfair actors) and personality differences (greedy and fair). In the ultimatum game, there are two players: a proposer and a receiver. The proposer is given an amount of money, such as \$1, and is tasked with offering a portion of the amount to the receiver. The receiver can either accept or reject the offer -- if the receiver accepts, the players divide the amount as proposed. If the receiver rejects, both players receive nothing. Economic theory dictates that a profit-maximizing proposer should offer only \$0.01 (the smallest nonzero amount) and keep \$0.99, and that the receiver should accept it because \$0.01 is more than the receiver would have otherwise. However, experiments with human subjects show that humans do not act in a purely ``rational'' manner; receivers will reject a low offer to punish proposers for offering an unfair split \parencite{KRAWCZYK2018283, vavra2018}. Moreover, proposers are aware of this, and thus strategically make offers that are closer to fair -- especially after multiple rounds of playing the game. 

We use the ultimatum game to test whether LLMs can simulate the strategic, social, and personality aspects of human reasoning.  
We extract human gameplay actions (offers and accept/reject decisions) from economics literature ~\parencite{HOUSER201419} and evaluate whether LLMs can simulate human behavior in the ultimatum game with 5 rounds. When the game is played for multiple rounds, both players have the opportunity to adjust their actions in response to the actions of the other player. We compare two LLM structures: a single LLM and a multi-agent system. We compare their abilities to (1) create realistic strategies, (2) adhere to created strategies, and (3) accurately model two different player personalities: greedy and fair. The single LLM structure involves prompting GPT4 directly, while the multi-agent system is adapted from recent literature ~\parencite{park2023generative}. 

In the single LLM structure, GPT4 is directly prompted to simulate the actions of both a proposer and receiver over five rounds of the ultimatum game. In the multi-agent system, each player is represented by a separate GPT4 agent. Each player is tasked with playing the ultimatum game with the other, with information such as personality hidden from the other agent. In both conditions, the LLM is tasked with creating a strategy based on a given personality and playing the game according to their personality and strategy. Prompting both structures to create strategies allows us to specifically analyze the consistency of LLM reasoning with human reasoning.

Our evaluation shows that the multi-agent system is significantly more accurate than using a single LLM to simulate strategic behavior in the ultimatum game. Over 40 simulations, the multi-agent system was consistent with human behavior 87.5\% of the time, while the single LLM was only consistent 50\% of the time. There are three causes of inconsistency between LLM simulations and human behavior: (1) a created strategy is incomplete, (2) a created strategy is inconsistent with the specified personality, or (3) a player deviates from the created strategy during game play. We find that over 90\% of issues in single LLM simulations are caused by the LLMs strategy -- i.e., their reasoning -- rather than the simulation of gameplay. Incomplete strategies and inconsistent personality strategies account for a roughly equal amount of errors. Only 1 out of 40 simulations has an error caused by a player not adhering to the created strategies. In the multi-agent system, the most common issue is strategies being inconsistent with personality, which accounts for more than 85\% of errors.

Based on these results from the ultimatum game, we believe multi-agent systems show potential to simulate plausible human behavior consistent with experimental evidence in more complex scenarios involving strategic reasoning. These systems can become a tool for decision makers in making plans, policies, and interfaces of which overall outcomes are influenced by reasoning at the individual level.

\section{Related Work}

\subsection{Human Reasoning in Ultimatum Games}
Experiments show that human subjects often reject low offers in the ultimatum game: 90\% of the time, receivers reject low offers of 10\% of the money \parencite{KRAWCZYK2018283}. Human subjects most commonly propose offers of 40\%--50\% of the money, with the receiver typically accepting ~\parencite{HOUSER201419}.
 
Introducing personality traits or multiple rounds into the ultimatum game has a demonstrable effect on player reasoning. Human proposers with ``selfish" personality traits make skewed offers ~\parencite{konigsteinmanfred}. Human receivers with ``fair" personality traits reject low offers to ``punish" proposers despite guaranteeing a worse outcome for themselves \parencite{vavra2018}.

\subsection{Prompting LLMs to Reason Improves Performance}

Previous work has shown that LLMs can be asked to create thought processes before acting, not only to enable researchers to follow their reasoning, but also to improve the accuracy of results.  Asking LLMs to think through intermediate steps improves arithmetic, symbolic, and logical reasoning \parencite{kojima2023large}. When prompted to explain intermediate reasoning, LLMs outperform human benchmarks on tasks in which standard prompting fails \parencite{suzgun2022challenging} and accuracy on grade-school math problems improves from 18\% to 57\% ~\parencite{DBLP:journals/corr/abs-2201-11903}. Progressive-hint prompting by a user improves the average accuracy of results by 20\% compared to standard prompting \parencite{zheng2023progressivehint}. Prompting GPT to create strategies before simulating the ultimatum game thus allows us to follow the LLM's reasoning and should improve accuracy of outcomes in our work.

\subsection{LLMs can Simulate Strategic Reasoning}

Prior research studying the degree to which GPT can simulate human strategic reasoning has yielded mixed results. GPT has been observed to under-perform compared to human benchmarks (55\% and 60\% accuracy) with Theory of Mind tasks \parencite{sap2023neural}, but a single LLM produces results very similar to human baselines in simulating the ultimatum game (three out of four measured offer thresholds from human studies fall on the LLM trendline, with the fourth deviating by less than 10\%) \parencite{aher2023using}. 

Thus, there is enough promise to study GPT's reasoning capabilities further in specific scenarios such as the ultimatum game. Previous work has suggested that LLMs can reason and negotiate like humans in various strategic scenarios \parencite{gandhi2023strategic}. LLMs have been observed to be able to make adjustments in reasoning in the middle of simulated hiring processes \parencite{horton2023large} and out-negotiate humans in an online Diplomacy league \parencite{bakhtin2022human}. GPT-4, OpenAI's newest LLM, specifically shows improved logical grounding and reasoning compared to its predecessors \parencite{bubeck2023sparks}, giving reason to further study its capabilities. 

\subsection{Multi-Agent System Reasoning}
Previous studies of GPT’s ability to simulate economic games have primarily used a single LLM, but multi-agent systems show more promise in simulating human reasoning, decision-making, and collaboration in social systems \parencite{li2023metaagents, Ghaffarzadegan_2024}. New multi-agent systems demonstrate emergent social behavior \parencite{chen2023agentverse, park2023generative} and the ability to simulate human reasoning and decision making in various contexts, including supreme court decisions \parencite{hamilton2023blind}, during epidemics \parencite{williams2023epidemic}, and the daily lives of inhabitants of a town ~\parencite{park2023generative}. Prior research involving economic games has demonstrated that multi-agent system results align with human trust behaviors and strategic behavior ~\parencite{guo2023gpt, xie2024large}, but unlike our work, this prior work does not prompt GPT agents to create strategies before playing.
\section{Experimental Set-Up}

To test the ability of LLMs to simulate strategic reasoning, we ran simulations of the five-round ultimatum game. We compared two different structures, a single LLM and a multi-agent system. We also tested the structures' abilities to model two personalities, greedy and fair. We ran 10 simulations for each personality pair, resulting in 40 simulations total. We selected greedy and fair personality types based on studies with human subjects: we expected differences in created strategies and  progression towards an equal split. For instance, we expected the initial offer in a simulation with a fair proposer and a fair receiver to be an even (\$0.50) or close-to-even split and to be accepted ~\parencite{HOUSER201419}. In contrast, we expected the initial offer in a simulation with a greedy proposer and fair receiver to be skewed in favor of the proposer ~\parencite{konigsteinmanfred} and to be rejected~\parencite{KRAWCZYK2018283, vavra2018}.

For all experiments, we used OpenAI’s GPT. We ran simulations with GPT-3.5 (gpt-3.5-turbo) and GPT-4 (gpt-4-1106-preview). GPT-4 has been demonstrated to interpret human concepts like equity \parencite{openai2023gpt4} and demonstrate improved reasoning abilities compared to GPT-3.5 \parencite{bubeck2023sparks}. However, GPT3.5 may be more accessible for policy makers and is thus worth testing. For both models, we set temperature and top P parameters to 0.5, and did not add frequency or presence penalties.  
 
\subsection{Research Questions}
We specifically address the following questions:

\bigskip
RQ1. Which LLM structure (single or multi-agent) more accurately simulates human-like actions in the five-round ultimatum game?

RQ2. Which LLM structure more accurately simulates reasoning based on personality?

RQ3. Which LLM structure more often creates robust strategies: both logically complete and consistent with personality?
\bigskip

Single LLMs and multi-agent systems have different advantages. A single LLM has full context for both players, potentially allowing more coherent narratives in simulation. Multi-agent systems lack global context, as agents can only exchange information through "speaking", but they more closely resemble inter-human communication and the anonymous conditions of the game.

\subsection{Single LLM and Multi-Agent System}
\subsubsection{Inputs}

For the single LLM, we prompt an LLM to create strategies for both players and then simulate the game. The prompt for two fair players is: ``Create a strategy for a \textit{fair} proposer and a \textit{fair} receiver in playing the ultimatum game five times with \$1. Once the strategies are created, simulate five rounds of the ultimatum game with the proposer and the receiver adhering to the previously outlined strategies." The prompts for other personality-pairings differ by only the italicized personality descriptors. From the LLM's response, we extract strategies and offers/responses for all five rounds.

For the multi-agent system, we adapt a previous architecture \parencite{park2023generative}. Each agent is specified with a name, public/private biographies, directives, and an initial plan. For our experiments, we name the agents "Proposer" or "Receiver", and set the initial plans as creating strategies. We specify player personalities (e.g., "Proposer is greedy.") in the private biography so that it is hidden from the other agent. The public biographies are blank, since players are not given any information about one another. Communication between agents in the architecture is not inherently turn-taking, so each agent is directed to wait for responses/offers before responding.

\subsubsection{Outputs}


\begin{figure}[t]
    \centering
	\includegraphics[trim={0.1cm 0.1cm 0.1cm 0.1cm}, clip,width=1\linewidth]{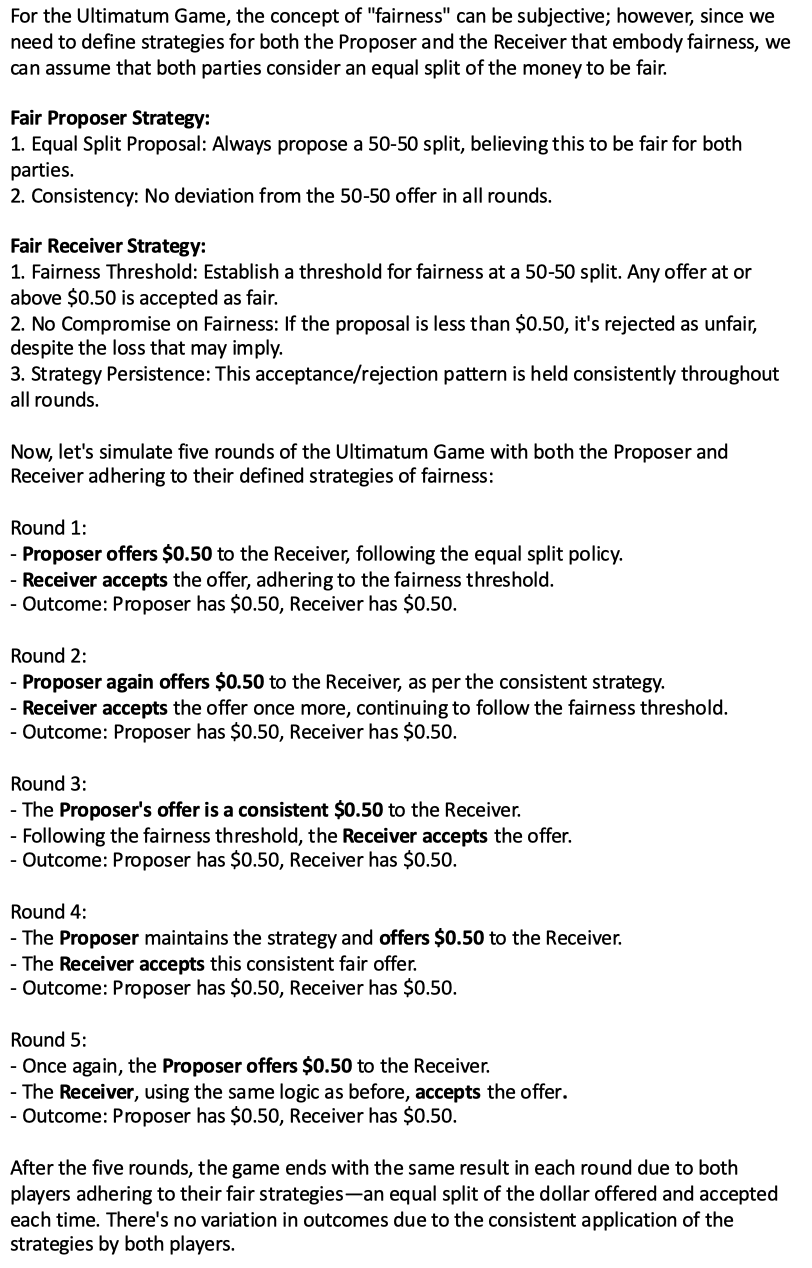}
	\caption{An output log from a SingleLLM simulation of two fair players playing five rounds of the ultimatum game. All text and indentation is from the LLM. The authors bolded some text to highlight strategy and gameplay actions.}
    \label{singleLLMsim}
\end{figure}

\begin{figure}[t]
    \centering
    \includegraphics[trim={0.1cm 0.1cm 0.1cm 0.1cm}, clip, width=1\linewidth]{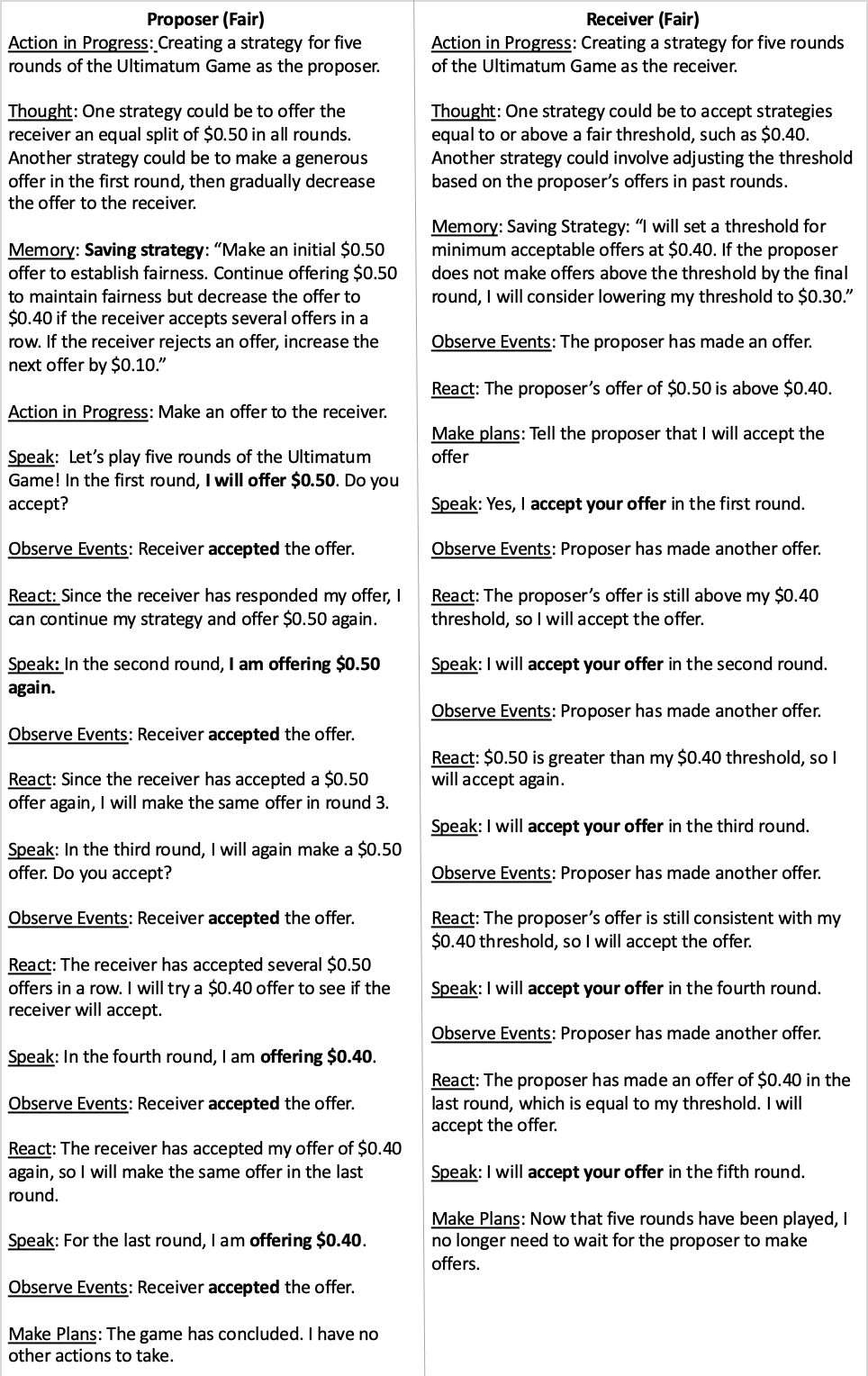}
    \caption{An output log from a Multi-Agent simulation of two fair players playing five rounds of the ultimatum game. All text is from the LLM; the labels (underlined) are provided by the architecture. The authors bolded some text to highlight strategy and gameplay actions.}
    \label{multiAgentsimTest}
\end{figure}

With the single LLM, outputs are displayed in one log. Player strategies typically involve an instruction for the first round’s offer/response and subsequent adjustments. Each round consists of three lines: the proposer’s offer, the receiver’s response, and the outcome. Figure~\ref{singleLLMsim} shows an example log with two fair players: the proposer strategizes to offer \$0.50, while the receiver uses \$0.50 as a fair threshold. This results in five accepted offers.

With the multi-agent system, outputs are displayed in two logs, one for each agent. Figure~\ref{multiAgentsimTest} shows example logs with two fair players. Each agent first creates a strategy. The proposer creates a strategy to offer \$0.50 to the receiver, and considers lowering this later. The receiver creates a strategy to reject offers below a \$0.40 threshold. The simulation results in \$0.50 offers in the first three rounds and \$0.40 offers in the last two rounds, resulting in five accepted offers.

We record the strategies of each player, and offers/outcome in each round.

\section{Evaluation}
\subsection{Evaluation of Gameplay}

Based on large-scale studies of human players \parencite{HOUSER201419, KRAWCZYK2018283}, we establish ranges of offers and answers for each personality type. Prior experiments with human studies show that fair proposers will offer equal or close to equal splits between the range of \$0.40 to \$0.50, with fair receivers typically accepting offers and greedy receivers typically rejecting. Meanwhile, greedy proposers offer initial splits heavily biased in their favor, typically above \$0.70, which is typically rejected by both a fair and greedy receiver.

We evaluate the initial offers of each simulation based on these criteria. In the first round, fair proposers are considered to act consistently with their personality if their offer is between \$0.40 and \$0.60, inclusive. Greedy proposers are considered to act consistently with their personality if their offer is biased in their favor, i.e., offering strictly less than \$0.50 to the receiver. Fair receivers are considered to act consistently with their personality if they reject offers that are less than \$0.40 and accept offers that are greater than or equal to \$0.40. Greedy receivers, however, are only considered to act consistently with their personality if they accept offers that are strictly greater than \$0.50; if a greedy receiver accepts any amount less than or equal to \$0.50, we consider the receiver to have not acted consistently with the greedy personality.

In subsequent rounds, we check if each player continues to act as per their created strategy as well as whether the taken action is consistent with results from human studies. Proposers are expected to continue making offers similar to the range of the initial offer if the receiver accepts, but if the receiver rejects, proposers are expected to increase their offers slightly \parencite{KRAWCZYK2018283}. Receivers are expected to accept offers as per their initial thresholds as well, but if gameplay progresses with no accepted offers, receivers are expected to lower their threshold, and potentially even discard it by the fifth round as there are no future rounds to influence offers.

\subsection{Evaluation of Strategies}
From the information collected from the LLM outputs, we evaluate strategies for three components: (1) the completeness of strategies, (2) the consistency of strategies with the specified personality trait, and (3) the adherence to the strategies in the following gameplay.

Strategies are considered complete if the player has a course of action for all possible states of the game. To be complete, a proposer’s strategy has to include an initial offer plan, and then a course of action for subsequent rounds based on whether the receiver accepts or rejects the previous offer. If the proposer’s strategy is incomplete, there can be issues with the proposer acting inappropriately when the receiver does not take the action for which the rest of the strategy is contingent on. Similarly, to be complete, a receiver’s strategy has to include a course of action for all five rounds for all possible offers between \$0.00 to \$1.00, typically specified via an acceptance threshold based on which the receiver acts.

For example, an incomplete strategy for a greedy proposer is as follows:

``Low-Ball Offers: The greedy proposer would aim to keep as much money as possible for themselves. They might start with a low offer to test the receiver's limit. Since we're dealing with \$1, the proposer may start by offering \$0.10 to the receiver.
\\
\\
Incremental Increase: If the offer is rejected, in subsequent rounds, they may increase the offer by a small increment, just enough to tempt the receiver to accept. For example, the proposer might increase the offer by \$0.05 each time.
\\
\\
Cut-Off Point: The proposer will have a cut-off point where they find it no longer worth to increase the offer because they would rather end up with nothing than give away more." 

The strategy does not account for the receiver accepting the first offer, potentially resulting in problematic gameplay from the proposer if this case is reached.

Strategies are consistent with the specified personality if the offers made (for the proposer) or rejected/accepted (for the receiver) are biased towards the player for greedy players and closer to an equal split for fair players. For example, a greedy proposer’s strategy should be to make low initial offers that are biased in the proposer’s favor, while a fair proposer’s strategy should be to make initial offers that are equal or close to equal.  Similarly, a greedy receiver’s strategy should be to only accept initial offers biased in the receiver’s favor, while a fair receiver’s strategy should be to accept initial offers that are equal or close to equal. In subsequent rounds, the strategy should be generally similar, although based on the actions of the other players, there may be concessions made by either player to reach agreements. For instance, even if offers are not biased in the favor of a greedy player, the strategy should also consider that something is better than nothing.

For example, a strategy inconsistent with personality for a greedy receiver is as follows: 

``Reject Low Offers: Initial minimum acceptance threshold is set high with a rejection of any offer below \$0.40. Accept all offers above \$0.40.
\\
\\
Willingness to Adjust: If offers remain low, be willing to gradually lower the acceptance threshold to ensure some gain.
\\
\\
Last Round Acceptance: On the final round, accept any non-zero offer, under the assumption that some gain is better than none, adjusting the minimum threshold to \$0.15." 

This strategy sets an acceptance threshold of \$0.40, which is lower than an equal split, and hence inconsistent with a greedy receiver whom would be expected to prefer offers that are biased in their favor (at least above \$0.50).

\section{Results}

\color{black}
We analyze the outputs of 40 simulations of the five-round ultimatum game each for 4 conditions: 
\begin{itemize}
\item multi-agent LLM system with GPT 3.5 (abbreviated ``MultiAgent-3.5")
\item multi-agent LLM system with GPT 4 (abbreviated ``MultiAgent-3.5")
\item a single LLM  with GPT 3.5 (abbreviated ``SingleLLM-3.5")
\item a single LLM with GPT 3.5 (abbreviated ``SingleLLM-4"). 
\end{itemize}
We report results for our three research questions. 
\color{black}

\bigskip
\noindent\textbf{RQ1: Which LLM structure (single or multi-agent) more accurately simulates human-like actions in the five-round ultimatum game?}

Our results show that the multi-agent systems yields actions consistent with human experimental data significantly more often than the single LLM. As shown in Table~\ref{outcome_numbers}, the best multi-agent structure was MultiAgent-4, which resulted in human-like actions in 87.5\% of simulations, while the best single LLM, SingleLLM-4, only resulted in human-like actions in 50\% of the simulations. A chi-square test shows this is statistically significant at the \(p < .01\) level: \(\chi^2(1, N = 80) = 13.091, p = .000297\).

An analysis of the errors shows that strategy creation was a bigger source of errors than gameplay mistakes for both structures. Table~\ref{error_explanations} shows the percentages of errors due to strategy, gameplay, or both for all four conditions. In both MultiAgent structures, strategy creation errors accounted for 100\% of errors in simulation, with there being no gameplay mistakes. In the SingleLLM-3.5 structure, 73.9\% of errors were in strategy creation, compared to only 39.1\% in gameplay (and 13.0\% having both types of errors). In the SingleLLM-4 structure, 100\% of errors involved an issue with strategy creation, with 25\% of errors also including gameplay mistakes. As shown in Table~\ref{error_explanations}, two-proportion z-tests revealed a statistically significant difference between the number of strategy creation errors and gameplay mistakes for all four conditions at a \(p < 0.05\) level.

\begin{table} 
\centering
\setlength{\tabcolsep}{3pt}
\begin{tabular}{||c| c||} 
 \hline
 Structure & Human Consistent Simulations (\%) \\ [0.5ex] 
 \hline\hline
 MultiAgent-3.5 & 82.5\% \\ 
 \hline
 \textbf{MultiAgent-4} & \textbf{87.5\%} \\
 \hline
 SingleLLM-3.5 & 42.5\% \\
 \hline
 SingleLLM-4 & 50.0\% \\
 \hline
\end{tabular}
\caption{\label{outcome_numbers}Percentage of simulations with human-like outcomes (RQ1). Most successful structure in bold.}
\end{table}

\begin{table*} 
\centering
\setlength{\tabcolsep}{5pt}
\begin{tabular}{|l|l l l l l|}
\hline
\textbf{Structure} & \textbf{Total Errors} & \textbf{Strategy Errors} & \textbf{Gameplay Errors} & \textbf{Both Errors} & \textbf{z-test} \\
\hline\hline
MultiAgent-3.5 & 7 & \textbf{100\%} (7/7) & 0\% (0/7) & 0\% (0/7) & \(z = 3.7417, p = .00018\)\\
\hline
MultiAgent-4 & 5 & \textbf{100\%} (5/5) & 0\% (0/5) & 0\% (0/5) & \(z = 3.1632, p = .00158\)\\
\hline
SingleLLM-3.5 & 23 & \textbf{73.9\%} (17/23) & 39.1\% (9/23) & 13.0\% (3/23) & \(z = 2.379, p = .017\)\\
\hline
SingleLLM-4 & 20 & \textbf{100\%} (20/20) & 25\% (5/20) & 25\% (5/20) & \(z = 4.899, p < .00001\) \\
\hline
\end{tabular}
\caption{\label{error_explanations}Number and percentage breakdown by type for errors in each structure (RQ1). In all structures, strategy errors are the most common source of issues - highest error source percentage in bold.}
\end{table*}

\bigskip
\noindent\textbf{RQ2: Which LLM structure more accurately simulates reasoning based on personality?  }

The experiments show that  MultiAgent-4  performed best at modeling the two personality types. MultiAgent-4 achieved human-like gameplay for all four personality pairs at least 80\% of the time (see Table \ref{outcome_by_personality}). Personality pairs are denoted as ``proposer personality"-``receiver personality". In contrast, SingleLLM-4 was inconsistent across personality pairs; it achieved human-like gameplay for 100\% of the Fair-Fair simulations, but only 10\% of the Greedy-Greedy conditions.

When analyzing gameplay for each of the personality pairs, we observe that the errors are not the same across the pairs. Fair-Fair has the best performance with SingleLLM-4, MultiAgent-3.5, and MultiAgent-4 all being 100\% consistent with human gameplay. 
The most errors occurred in simulations of the Greedy-Greedy personality pairing, with MultiAgent-4 performing the best with 80\% of simulations being consistent with human gameplay. MultiAgent-3.5, SingleLLM-3.5, and SingleLLM-4 were consistent with human gameplay in 70\%, 60\%, and 10\% of Greedy-Greedy simulations respectively. The Fair-Greedy and Greedy-Fair conditions were somewhere in between: both single LLMs had middling scores (30-50\%).

\begin{table*} 
\centering
\begin{tabular}{|l|c c c c|}
\hline
\textbf{Structure} & \textbf{Fair-Fair} & \textbf{Fair-Greedy} & \textbf{Greedy-Fair} & \textbf{Greedy-Greedy} \\
\hline\hline
MultiAgent-3.5 & \textbf{100\%} & \textbf{80\%} & 80\% & 70\% \\ 
\hline
\textbf{MultiAgent-4} & \textbf{100\%} & \textbf{80\%} & \textbf{90\%} & \textbf{80\%} \\
\hline
SingleLLM-3.5 & 30\% & 50\% & 30\% & 60\% \\
\hline
SingleLLM-4 & \textbf{100\%} & 40\% & 50\% & 10\% \\
\hline
\end{tabular}
\caption{\label{outcome_by_personality}Percentage of simulations with human-like outcomes (RQ2). Most successful structure(s) in bold.}
\end{table*}


\bigskip
\textbf{RQ3. Which LLM structure more often creates robust strategies: both logically complete and consistent with personality?}

The multi-agent systems create robust strategies at a higher rate than single LLMs (see Table \ref{strategies_eval}). MultiAgent-4 creates complete and personality-consistent strategies for both players in 87.5\% of simulations. MultiAgent-3.5 performs slightly worse, creating complete and personality-consistent strategies for both players in 80\% of simulations. SingleLLM-3.5 and SingleLLM-4 create complete and personality-consistent strategies in 55\% and 47.5\% of simulations respectively.

We find that the MultiAgent-4 structure performs better in creating complete and personality-consistent strategies than the best-performing SingleLLM structure (SingleLLM-3.5). A chi-square test shows this is statistically significant at the \(p < .01\) level: \(\chi^2(1, N = 40) = 10.3127, p = .001321\).

To analyze the source of these errors, we analyze the robustness of proposer strategies and receiver strategies separately. 
Table \ref{proposer_strategies} shows that the problem with proposer strategies is always incompleteness. Proposers have no errors with personality consistency across all four structures. Conversely, Table \ref{receiver_strategies} shows that the problem with receiver strategies with issues are almost always inconsistent with personality. 
Across all conditions, there was only one incomplete receiver strategy.

\begin{table*}  
\centering
\begin{tabular}{|l|>{\centering\arraybackslash}p{3cm}|>{\centering\arraybackslash}p{3cm}|>{\centering\arraybackslash}p{3cm}|}
\hline
\textbf{Structure} & \textbf{\% Strategies Complete} & \textbf{\% Strategies Consistent with Personality} & \textbf{\% Strategies Complete \& Consistent} \\
\hline\hline
MultiAgent-3.5 & 90\% & 85\% & 80\% \\
\hline
MultiAgent-4 & \textbf{95\%} & \textbf{87.5\%} & \textbf{87.5\%} \\
\hline
SingleLLM-3.5 & 65\% & 80\% & 55\% \\
\hline
SingleLLM-4 & 55\% & 60\% & 47.5\% \\
\hline
\end{tabular}
\caption{\label{strategies_eval}Percentage of simulations in which both strategies are complete, consistent, and both (RQ3). Most successful structure in bold.}
\end{table*}

\begin{table*} 
\centering
\begin{tabular}{|l|c c|}
\hline
\textbf{Structure} & \multicolumn{1}{|p{4.5cm}|}{\centering \textbf{Proposer: \\ \% Strategies Complete}} & \multicolumn{1}{p{4.5cm}|}{\centering \textbf{Proposer: \\ \% Strategies Consistent with Personality}} \\
\hline\hline
MultiAgent-3.5 & \color{red}92.5\%\color{black} & 100\% \\
\hline
MultiAgent-4 & \color{red}95\%\color{black} & 100\% \\
\hline
SingleLLM-3.5 & \color{red}67.5\%\color{black} & 100\% \\
\hline
SingleLLM-4 & \color{red}52.5\%\color{black} & 100\% \\
\hline
\end{tabular}
\caption{\label{proposer_strategies}Percentage of proposer strategies that are complete, consistent, and both (RQ3). Red indicates the presence of errors.}
\end{table*}
\begin{table*} 
\centering
\begin{tabular}{|l|c c|}
\hline
\textbf{Structure} & \multicolumn{1}{|p{4.5cm}|}{\centering \textbf{Receiver: \\ \% Strategies Complete}} & \multicolumn{1}{p{4.5cm}|}{\centering \textbf{Receiver: \\ \% Strategies Consistent with Personality}} \\
\hline\hline
MultiAgent-3.5 & \color{red}97.5\%\color{black} & \color{red}85\%\color{black} \\
\hline
MultiAgent-4 & 100\% & \color{red}87.5\%\color{black} \\
\hline
SingleLLM-3.5 & 100\% & \color{red}80\%\color{black} \\
\hline
SingleLLM-4 & 100\% & \color{red}60\%\color{black} \\
\hline
\end{tabular}
\caption{\label{receiver_strategies}Percentage of receiver strategies that are complete, consistent, and both (RQ3). Red indicates the presence of errors.}
\end{table*}

\section{Discussion}
\subsection{Why are multi-agent systems better at strategic simulation?}

We found that multi-agent systems show greater promise than single LLMs for simulating strategic human reasoning. Multi-agent systems showed relatively high consistency with human behavior (87.5\%), simulated all personality pairings well (80\%-100\%), were generally able to produce complete (95\%) and consistent (87.5\%) reasoning, and adhere to the strategies in gameplay (100\%). In contrast, single agent LLMs were only 50\% consistent with human behavior, with 90\% of the errors coming poor strategies. This makes single LLMs less than ideal as a simulation tool.

Single LLM simulations most often fail because strategies are incomplete. The best performing single LLM only produced complete stratgies 65\% of the time. In comparison, Multi-Agent systems both had excellent strategy completion rates (90\% and 95\%). Seemingly asking a single LLM to come up with two strategies at once is sufficiently difficult that it creates incompleteness errors - it ``forgets" to think through all the cases of each personalities' strategy. This deficiency in the single LLMs ability to reason is likely the root cause of its problems. 

\subsection{LLM Simulations for Decision Makers}
Decision makers and policy-makers need to consider all of the ways in which individuals may react in response to new programs and policies in order to foresee potential consequences.
Strategic reasoning is especially important to simulate in policy design and security settings. Will greedy, malicious, lazy, or confused people break the system, intentionally or unintentionally? How will proposed solutions to unforeseen consequences fare? Thinking through all of these possibilities can be mentally demanding, but we propose that multi-agent systems have the potential to be an interactive tool to help designers explore a space of action consistent with human reasoning and take into account complexities like personality, ``irrationality”, and strategic thinking. 

Multi-agent systems can scale to handle hundreds of agents interacting within the system. This can allow for testing dozens of personality types, beyond just greedy and fair, in future work. Additionally, it can test full societies with different ratios of personality types. A society where the entire population is greedy may not survive, but a society with only 10-25\% greedy people may thrive because there are enough fair people to uphold the system.

We believe that simulations can be a tool that decision-makers can use quickly and easily. Simulations may not provide full solutions, but they can help decision-makers foresee how different types of people will reason and react. Currently, simulation tools are non-trivial to get running and extract results from. However, a future goal for the research community is to make such tools easy and accessible to use. 

\subsection{Limitations}

This paper studies human strategic reasoning with the ultimatum game as a case study. For larger examples and more complex scenarios, LLMs may not perform as well as they do in the ultimatum game. This version of the ultimatum game does not challenge the LLM's context window, output constraints, or attention mechanism. Further investigations should test ultimatum game variants with more rounds and players. We expect multi-agent systems to be good at this, but this should be tested in future work, perhaps on variants of the ultimatum game such as 
the competitive ultimatum game where multiple proposers make offers and receivers must pick among them. 

The ultimatum game might be too popular to be used as a test for generalized human behavior. 
LLMs are trained to make predictions based on their large text corpus. GPT may have examples of strategies and gameplay to draw from. Thus, it might not be performing strategic behavior that can be generalized to other scenarios, since it could just be recreating examples it has seen. However, this is unlikely because the single LLM performs poorly, with only the multi-agent system starts to get promising results. If the LLM were purely parroting back past examples, we would expect a single LLM to excel. Furthermore,  there is reason to be optimistic that LLMs have such a broad knowledge base that very little is truly new to them. Either way, future work should further explore how an LLM would be able to simulate strategic human behavior in novel scenarios. 

It is an additional challenge to simulate human behavior for truly unprecedented events with no history to draw from. This might include new technologies like AI in the workforce or advances in security. Without explicit data to draw from, LLMs would have to reason from first principles, or draw inferences from past events like previous emergencies or innovations and adjust them to modern times. It could be possible for an LLM to rely on social science theories of human behavior to base simulations on.  LLMs have shown a surprising ability to reason, rather than just recall information. In addition, even if they can’t reason completely about novel events, they can still be useful to designers in covering the less novel aspects of a complex situation as it evolves. This is a fertile and important area for researchers to explore in future work.

\section{Conclusion}
Based on our experiments with single LLMs and multi-agent systems, we conclude that multi-agent systems show great potential for simulating strategic behavior consistent with human gameplay. We compare LLMs playing the ultimatum game over 5 rounds and see that multi-agent systems achieve gameplay consistent with human experimental data in 85\% of simulations, while single LLMs achieve gameplay consistent with human data in only 50\% of simulations. Surprisingly, when the single LLMs make errors, 100\% had strategy creation issues, with 25\% also having gameplay issues.

Based on the strengths of multi-agent LLMs systems to create and execute strategic thinking and behavior, we believe these systems can become a tool for policy designers to think through the behavior of agents with different personalities, who are all trying to strategically navigate a system to achieve a personal outcome. This type of thinking is immensely difficult for people, and LLM-based simulations can aid this cognitive process.

\printbibliography

@book{ariely_2008,
	address = {New York, NY},
	title = {Predictably Irrational: The Hidden Forces That Shape Our Decisions},
	language = {English},
	publisher = {Harper},
	author = {Ariely, Dan},
	year = {2008}
}

@book{kahneman_thinking_2012,
	address = {London},
	title = {Thinking, fast and slow},
	language = {English},
	publisher = {Penguin},
	author = {Kahneman, Daniel},
	year = {2012}
}

@incollection{mccrae2008five,
  title={The five-factor theory of personality},
  author={McCrae, Robert R and Costa, Paul T Jr},
  booktitle={Handbook of personality: Theory and research},
  editor={John, Oliver P and Robins, Richard W and Pervin, Lawrence A},
  edition={3},
  pages={159--181},
  year={2008},
  publisher={The Guilford Press}
}

@article{KIDD2013109,
title = {Rational snacking: Young children’s decision-making on the marshmallow task is moderated by beliefs about environmental reliability},
journal = {Cognition},
volume = {126},
number = {1},
pages = {109-114},
year = {2013},
issn = {0010-0277},
doi = {https://doi.org/10.1016/j.cognition.2012.08.004},
url = {https://www.sciencedirect.com/science/article/pii/S0010027712001849},
author = {Celeste Kidd and Holly Palmeri and Richard N. Aslin}
}

@book{mullainathan2013scarcity,
  title={Scarcity: Why having too little means so much},
  author={Mullainathan, Sendhil and Shafir, Eldar},
  year={2013},
  publisher={Times Books/Henry Holt and Co}
}

@misc{hamilton2023blind,
      title={Blind Judgement: Agent-Based Supreme Court Modelling With GPT}, 
      author={Sil Hamilton},
      year={2023},
      eprint={2301.05327},
      archivePrefix={arXiv},
      primaryClass={cs.CL}
}

@misc{park2023generative,
      title={Generative Agents: Interactive Simulacra of Human Behavior}, 
      author={Joon Sung Park and Joseph C. O'Brien and Carrie J. Cai},
      year={2023},
      eprint={2304.03442},
      archivePrefix={arXiv},
      primaryClass={cs.HC}
}

@misc{aher2023using,
      title={Using Large Language Models to Simulate Multiple Humans and Replicate Human Subject Studies}, 
      author={Gati Aher and Rosa I. Arriaga and Adam Tauman Kalai},
      year={2023},
      eprint={2208.10264},
      archivePrefix={arXiv},
      primaryClass={cs.CL}
}

@incollection{KRAWCZYK2018283,
title = {Chapter 12 - Social Cognition: Reasoning With Others},
editor = {Daniel C. Krawczyk},
booktitle = {Reasoning},
publisher = {Academic Press},
pages = {283-311},
year = {2018},
isbn = {978-0-12-809285-9},
doi = {https://doi.org/10.1016/B978-0-12-809285-9.00012-0},
url = {https://www.sciencedirect.com/science/article/pii/B9780128092859000120},
author = {Daniel C. Krawczyk}
}

@incollection{HOUSER201419,
title = {Chapter 2 - Experimental Economics and Experimental Game Theory},
editor = {Paul W. Glimcher and Ernst Fehr},
booktitle = {Neuroeconomics (Second Edition)},
publisher = {Academic Press},
edition = {Second Edition},
address = {San Diego},
pages = {19-34},
year = {2014},
isbn = {978-0-12-416008-8},
doi = {https://doi.org/10.1016/B978-0-12-416008-8.00002-4},
url = {https://www.sciencedirect.com/science/article/pii/B9780124160088000024},
author = {Daniel Houser and Kevin McCabe}
}

@article{konigsteinmanfred,
author = {Königstein, Manfred},
year = {2001},
month = {10},
pages = {S53 - S70},
title = {Personality influences on Ultimatum Game bargaining decisions},
volume = {15},
journal = {European Journal of Personality},
doi = {10.1002/per.424}
}

@article{DBLP:journals/corr/abs-2201-11903,
  author       = {Jason Wei and
                  Xuezhi Wang and
                  Dale Schuurmans},
  title        = {Chain of Thought Prompting Elicits Reasoning in Large Language Models},
  journal      = {CoRR},
  volume       = {abs/2201.11903},
  year         = {2022},
  url          = {https://arxiv.org/abs/2201.11903},
  eprinttype    = {arXiv},
  eprint       = {2201.11903},
  bibsource    = {dblp computer science bibliography, https://dblp.org}
}

@misc{zheng2023progressivehint,
      title={Progressive-Hint Prompting Improves Reasoning in Large Language Models}, 
      author={Chuanyang Zheng and Zhengying Liu and Enze Xie and Zhenguo Li and Yu Li},
      year={2023},
      eprint={2304.09797},
      archivePrefix={arXiv},
      primaryClass={cs.CL}
}

@misc{horton2023large,
      title={Large Language Models as Simulated Economic Agents: What Can We Learn from Homo Silicus?}, 
      author={John J. Horton},
      year={2023},
      eprint={2301.07543},
      archivePrefix={arXiv},
      primaryClass={econ.GN}
}

@misc{guo2023gpt,
      title={GPT in Game Theory Experiments}, 
      author={Fulin Guo},
      year={2023},
      eprint={2305.05516},
      archivePrefix={arXiv},
      primaryClass={econ.GN}
}

@misc{openai2023gpt4,
      title={GPT-4 Technical Report}, 
      author={OpenAI},
      year={2023},
      eprint={2303.08774},
      archivePrefix={arXiv},
      primaryClass={cs.CL}
}

@article{ThreeDWorld,
  author       = {Chuang Gan and
                  Jeremy Schwartz and
                  Seth Alter},
  title        = {ThreeDWorld: {A} Platform for Interactive Multi-Modal Physical Simulation},
  journal      = {CoRR},
  volume       = {abs/2007.04954},
  year         = {2020},
  url          = {https://arxiv.org/abs/2007.04954},
  eprinttype    = {arXiv},
  eprint       = {2007.04954}
}

@misc{li2023metaagents,
      title={MetaAgents: Simulating Interactions of Human Behaviors for LLM-based Task-oriented Coordination via Collaborative Generative Agents}, 
      author={Yuan Li and Yixuan Zhang and Lichao Sun},
      year={2023},
      eprint={2310.06500},
      archivePrefix={arXiv},
      primaryClass={cs.AI}
}

@article{vavra2018,
    author={Vavra, Peter and Chang, Luke J. and Sanfey, Alan G.},   
	title={Expectations in the Ultimatum Game: Distinct Effects of Mean and Variance of Expected Offers},      
    journal={Frontiers in Psychology},      
    volume={9},           
	year={2018},      
    url={https://www.frontiersin.org/articles/10.3389/fpsyg.2018.00992}
}

@misc{suzgun2022challenging,
      title={Challenging BIG-Bench Tasks and Whether Chain-of-Thought Can Solve Them}, 
      author={Mirac Suzgun and Nathan Scales and Nathanael Schärli},
      year={2022},
      eprint={2210.09261},
      archivePrefix={arXiv},
      primaryClass={id='cs.CL' full_name='Computation and Language' is_active=True alt_name='cmp-lg' in_archive='cs' is_general=False description='Covers natural language processing. Roughly includes material in ACM Subject Class I.2.7. Note that work on artificial languages (programming languages, logics, formal systems) that does not explicitly address natural-language issues broadly construed (natural-language processing, computational linguistics, speech, text retrieval, etc.) is not appropriate for this area.'}
}

@misc{kojima2023large,
      title={Large Language Models are Zero-Shot Reasoners}, 
      author={Takeshi Kojima and Shixiang Shane Gu and Machel Reid and Yutaka Matsuo and Yusuke Iwasawa},
      year={2023},
      eprint={2205.11916},
      archivePrefix={arXiv},
      primaryClass={id='cs.CL' full_name='Computation and Language' is_active=True alt_name='cmp-lg' in_archive='cs' is_general=False description='Covers natural language processing. Roughly includes material in ACM Subject Class I.2.7. Note that work on artificial languages (programming languages, logics, formal systems) that does not explicitly address natural-language issues broadly construed (natural-language processing, computational linguistics, speech, text retrieval, etc.) is not appropriate for this area.'}
}

@misc{sap2023neural,
      title={Neural Theory-of-Mind? On the Limits of Social Intelligence in Large LMs}, 
      author={Maarten Sap and Ronan LeBras and Daniel Fried and Yejin Choi},
      year={2023},
      eprint={2210.13312},
      archivePrefix={arXiv},
      primaryClass={id='cs.CL' full_name='Computation and Language' is_active=True alt_name='cmp-lg' in_archive='cs' is_general=False description='Covers natural language processing. Roughly includes material in ACM Subject Class I.2.7. Note that work on artificial languages (programming languages, logics, formal systems) that does not explicitly address natural-language issues broadly construed (natural-language processing, computational linguistics, speech, text retrieval, etc.) is not appropriate for this area.'}
}

@article{bakhtin2022human,
  title={Human-level play in the game of \textit{diplomacy} by combining language models with strategic reasoning},
  author={Bakhtin, Anton and Brown, Noam and Dinan, Emily},
  journal={Science},
  volume={378},
  number={6624},
  pages={1067--1074},
  year={2022}
}

@article{gandhi2023strategic,
  title={Strategic Reasoning with Language Models},
  author={Gandhi, Kanishk and Sadigh, Dorsa and Goodman, Noah D.},
  journal={arXiv preprint arXiv:2301.12345},  % Replace with actual arXiv number if available
  year={2023}
}

@article{bubeck2023sparks,
  title={Sparks of Artificial General Intelligence: Early experiments with GPT-4},
  author={Bubeck, S{\'e}bastien and Chandrasekaran, Varun and Eldan, Ronen and Elhage},
  journal={arXiv preprint arXiv:2303.12712},
  year={2023}
}

@article{Ghaffarzadegan_2024,
   title={Generative agent‐based modeling: an introduction and tutorial},
   volume={40},
   ISSN={1099-1727},
   url={http://dx.doi.org/10.1002/sdr.1761},
   DOI={10.1002/sdr.1761},
   number={1},
   journal={System Dynamics Review},
   publisher={Wiley},
   author={Ghaffarzadegan, Navid and Majumdar, Aritra and Williams, Ross},
   year={2024},
   month=jan }

@misc{williams2023epidemic,
      title={Epidemic Modeling with Generative Agents}, 
      author={Ross Williams and Niyousha Hosseinichimeh and Aritra Majumdar},
      year={2023},
      eprint={2307.04986},
      archivePrefix={arXiv},
      primaryClass={id='cs.AI' full_name='Artificial Intelligence' is_active=True alt_name=None in_archive='cs' is_general=False description='Covers all areas of AI except Vision, Robotics, Machine Learning, Multiagent Systems, and Computation and Language (Natural Language Processing), which have separate subject areas. In particular, includes Expert Systems, Theorem Proving (although this may overlap with Logic in Computer Science), Knowledge Representation, Planning, and Uncertainty in AI. Roughly includes material in ACM Subject Classes I.2.0, I.2.1, I.2.3, I.2.4, I.2.8, and I.2.11.'}
}

@misc{chen2023agentverse,
      title={AgentVerse: Facilitating Multi-Agent Collaboration and Exploring Emergent Behaviors}, 
      author={Weize Chen and Yusheng Su and Jingwei Zuo and Cheng Yang},
      year={2023},
      eprint={2308.10848},
      archivePrefix={arXiv},
      primaryClass={id='cs.CL' full_name='Computation and Language' is_active=True alt_name='cmp-lg' in_archive='cs' is_general=False description='Covers natural language processing. Roughly includes material in ACM Subject Class I.2.7. Note that work on artificial languages (programming languages, logics, formal systems) that does not explicitly address natural-language issues broadly construed (natural-language processing, computational linguistics, speech, text retrieval, etc.) is not appropriate for this area.'}
}

@misc{xie2024large,
      title={Can Large Language Model Agents Simulate Human Trust Behaviors?}, 
      author={Chengxing Xie and Canyu Chen and Feiran Jia},
      year={2024},
      eprint={2402.04559},
      archivePrefix={arXiv},
      primaryClass={id='cs.AI' full_name='Artificial Intelligence' is_active=True alt_name=None in_archive='cs' is_general=False description='Covers all areas of AI except Vision, Robotics, Machine Learning, Multiagent Systems, and Computation and Language (Natural Language Processing), which have separate subject areas. In particular, includes Expert Systems, Theorem Proving (although this may overlap with Logic in Computer Science), Knowledge Representation, Planning, and Uncertainty in AI. Roughly includes material in ACM Subject Classes I.2.0, I.2.1, I.2.3, I.2.4, I.2.8, and I.2.11.'}
}

@inproceedings{herr2019bluecollar,
  author = {Herr, Dominik and Grund, Sebastian and Ertl, Thomas},
  title = {BlueCollar: Optimizing Worker Paths on Factory Shop Floors with Visual Analytics},
  booktitle = {Proceedings of the 52nd Hawaii International Conference on System Sciences},
  year = {2019},
  url = {https://scholarspace.manoa.hawaii.edu/items/5cd6f86c-7c56-4f62-9071-f3b4c9620533},
  note = {Accessed: 2024-06-11}
}

@inproceedings{kasik2019toward,
  author = {Kasik, David and Dill, John},
  title = {Toward Technology Transfer Evaluation Criteria},
  booktitle = {Proceedings of the 52nd Hawaii International Conference on System Sciences},
  year = {2019},
  note = {Presented on 2019-01-08}
}

@inproceedings{zhao2020route,
  author = {Zhao, Jieqiong and Karimzadeh, Morteza and Xu, Hanye},
  title = {Route Packing: Geospatially-Accurate Visualization of Route Networks},
  booktitle = {Proceedings of the 53rd Hawaii International Conference on System Sciences},
  year = {2020},
  note = {Presented on 2020-01-07}
}

\end{document}